\documentclass[12pt]{iopart}
\usepackage{iopams}
\usepackage[dvips]{graphicx}

\begin{document}

\title[Preparation of Schr\"{o}dinger cat states ...]{Preparation of Schr\"{o}dinger cat states of a cavity field via coupling to a superconducting charge qubit}


\author{Dagoberto S. Freitas$^{1,2}$\footnote{Author to whom any
correspondence should be addressed.\\ Permanent address:
Departamento de F\'\i sica, Universidade Estadual de Feira de
Santana, 44036-900, Feira de Santana, BA, Brazil.} and M. C.
Nemes$^1$}

\address{$ˆ1$ Departamento de F\'\i sica,
Instituto de Ci\^encias Exatas, Universidade Federal de Minas
Gerais, 30123-970, Belo Horizonte, MG, Brazil}
\address{$^2$ Departamento de F\'\i
sica, Universidade Estadual de Feira de Santana, 44036-900, Feira de
Santana, BA, Brazil}

\ead{dfreitas@uefs.br}


\begin{abstract}
We extend the approach of Ref. \cite{yu04} for preparing
superposition states of a cavity field interacting with a
superconducting charge qubit. We study effects of the nonlinearity
on the creation of such states. We show that the main contribution
of nonlinear effects is to shorten the time necessary to build the
superposition.
\end{abstract}

\pacs{42.50.Dv, 42.50.Ct, 74.50.+r}

\maketitle

\section{Introduction}
The generation of nonclassical states of a radiation field has
become increasingly important in the past years given its various
applications in quantum communication. The feasibility of generating
such nonclassical states has been established in several branches of
physics such as cavity electrodynamics, trapped ions, quantum dots,
atoms inside cavities and so on \cite{brune96, monroe96}. In this
contribution we discuss the issue in the context of a
superconducting qubit in microcavity. It has been recently proposed
a way to engineer quantum states using a SQUID charge qubit inside a
cavity \cite{yu04, yu05} with a controllable interaction between the
cavity field and the charge qubit. The key ingredients are a tunable
gate voltage and a classical magnetic field applied to SQUID. In
Ref. \cite{yu04} a model is proposed including these ingredients and
using some adequate approximations which allows for the
linearization of the interaction and nonclassical states of the
field are generated. In Ref. \cite{yu05} the same model is used for
the preparation of macroscopic quantum superposition states of a
cavity field via coupling to a superconducting charge qubit.

We show that the essential contribution of nonlinear interaction is
to shorten the time necessary to build the quantum state. Since
decoherence is known to affect quantum effects uninterruptedly, they
are working even when the quantum state is being formed. This has
been studied and quantified in the context of cavity QED where it is
shown that the more ``quantum" is the superposition more rapid are
the environmental effects during the process of creating the quantum
state \cite{ze02}. It is therefore interesting to envisage processes
through which quantum superpositions are generated as fast as
possible.

\section{The model}
We consider a system constituted by a SQUID type superconducting box
with $n_{c}$ excess Cooper-pair charges connected to a
superconducting loop via two identical Josephson junctions having
capacitors $C_{J}$ and coupling energies $E_{J}$. An external
control voltage $V_{g}$ couples to the box via a capacitor $C_{g}$.
We also assume that the system operates in a regime consistent with
most experiments involving charge qubits, in which only Cooper pairs
coherently tunnel in the superconducting junctions. Therefore the
system Hamiltonian may be written as \cite{yu01}
\begin{equation}
H_{qb} = 4E_{ch}(n_{c}-n_{g})^{2}-2E_{j}\cos(\frac{\pi
\Phi_{X}}{\Phi_{0}})\cos(\Theta),\label{Hqb}
\end{equation}
where $E_{ch} = e^{2}/2(C_{g} +2C_{J})$ is the single-electron
charging energy, $n_{g} = C_{g}V_{g}/2e$ is the dimensionless gate
charge (controlled by $V_{g}$), $\Phi_{X}$ is the total flux through
the SQUID loop and $\Phi_{0}$ the flux quantum. By adjusting the
flux through the superconducting loop, one may control the Josephson
coupling energy as well as switch on and off the qubit-field
interaction. The phase $\Theta = (\phi_{1}+\phi_{2})/2$ is the
quantum-mechanical conjugate of the number operator $n_{c}$ of the
Cooper pairs in the box, where $\phi_{i}$ (i = 1, 2) is the phase
difference for each junction. The superconducting box is assumed to
be working in the charging regime and the superconducting energy gap
$\Delta$ is considered to be the largest energy involved. Moreover,
the temperature $T$ is low enough so that the conditions  $\Delta
\gg E_{ch} \gg E_{J} \gg k_{B}T$, where $k_{B}$ is the Boltzmann
constant. The superconducting box then becomes a two-level system
with states $|g\rangle$ (for $n_{c} = 0$) and $|e\rangle$ (for
$n_{c} = 1$) given that the gate voltage is near a degeneracy point
($n_{g} = 1/2$) \cite{yu01} and the quasi-particle excitation is
completely suppressed \cite{averin}.

If the circuit is placed within a single-mode microwave
superconducting cavity, the qubit can be coupled to both a classical
magnetic field (generates a flux $\Phi_{c}$) and the quantized
cavity field (generates a flux $\Phi_{q} =\eta a +\eta^{*}
a^{\dag}$, with $a$ and $a^{\dag}$ the annihilation and creation
operators), being the total flux through the SQUID given by
$\Phi_{X} = \Phi_{c} + \Phi_{q}$ \cite{you03,saidi02,zhu03}. The
parameter $\eta$ is related to the mode function of the cavity
field. The system Hamiltonian will then read
\begin{equation}
H = \hbar\omega a^{\dag}a + E_{z}\sigma_{z} - E_{J}(\sigma_{+} +
\sigma_{-})\cos\big(\gamma I + \beta a +\beta^{*}
a^{\dag}\big),\label{H}
\end{equation}
where we have defined the parameters $\gamma = \pi\Phi_{c}/\Phi_{0}$
and $\beta = \pi\eta/\Phi_{0}$. The first term corresponds to the
free cavity field with frequency $\omega = 4E_{ch}/\hbar$ and the
second one to the qubit having energy $E_{z} = -2E_{ch}(1 - 2n_{g})$
with $\sigma_z = |e\rangle\langle e| - |g\rangle\langle g|$. The
third term is the (nonlinear) photon-qubit interaction term which
may be controlled by the classical flux $\Phi_{c}$. In general the
Hamiltonian in equation (\ref{H}) is linearized under some kind of
assumption. In Ref. \cite{yu04}, for instance, the authors
decomposed the cosine in Eq.(\ref{H}) and expanded the terms
$\sin[\pi(\eta\, a+H.c.)/\Phi_{0}]$ and $\cos[\pi(\eta\,
a+H.c.)/\Phi_{0}]$ as power series in $a \,(a^{\dagger})$. In this
way, if the limit $|\beta|\ll 1$ is taken, only single-photon
transition terms in the expansion are kept, and  a Jaynes-Cummings
type Hamiltonian (JCM) is then obtained. Here, in contrast to that,
we adopt a similar technique to the one presented in reference
\cite{moya} obtaining a JCM Hamiltonian valid for any value of
$|\beta|$.

\section{Dynamics of the system}
The key idea of the method proposed in Ref. \cite{moya} which allows
for the inclusion of nonlinear effects is the following: a unitary
transformation is constructed in a way that diagonalizes the
Hamiltonian leading it to a much simpler form. The nonlinear effects
are therefore held in the transformation affecting directly the time
evolution of the system in a tractable manner.

We first apply a unitary transformation to the full Hamiltonian
given by (\ref{H}) and make approximations afterwards. By applying
the unitary transformation \cite{moya}
\begin{eqnarray}
T &=& \frac{1}{\sqrt{2}} \Big\{-\frac{1}{2} \Big[
 D^{\dag}\Big(\alpha,\gamma \Big) -  D\Big(\alpha,\gamma \Big)
\Big]I-\frac{1}{2} \Big[
 D^{\dag}\Big(\alpha,\gamma\Big) + D\Big(\alpha,\gamma
\Big) \Big] \sigma_z\nonumber\\
&+& D\Big(\alpha,\gamma \Big)\sigma_{+} + D^{\dag}\Big(\alpha,\gamma
\Big) \sigma_{-} \Big\} \label{T}
\end{eqnarray}
to the Hamiltonian in equation (\ref{H}), with
$D(\alpha,\gamma)=D(\alpha)e^{i\frac{\gamma}{2}}$ where $D(\alpha)
=\exp[(\alpha a^{\dag} -\alpha^{*}a)]$ is the Glauber's displacement
operator, with $\alpha = i\beta^{*}/2$, we obtain the following
transformed Hamiltonian
\begin{eqnarray}
H_{T}  & \equiv &  THT^{\dagger}\nonumber\\
& = & \hbar\omega
a^{\dag}a+\frac{E_{J}}{2}\sigma_{z}+i\frac{\hbar}{2}\big[\omega\big(\beta
a- \beta^{*}a^{\dag}\big) +
2i\frac{E_{z}}{\hbar}\big]\sigma_{x}\nonumber\\
& + & \frac{E_{J}}{2}\cos\big[2\big(\beta a+
\beta^{*}a^{\dag}\big)+2\gamma
\big]\sigma_{z}\nonumber\\
&-& i\frac{E_{J}}{2}\sin\big[2\big(\beta a+
\beta^{*}a^{\dag}\big)+2\gamma
\big]\big(\sigma_{+}-\sigma_{-}\big)+|\frac{\beta}{2}|^{2}.\label{HT0}
\end{eqnarray}
This result holds for any value of the parameter $\beta$. For the
regime in which $\hbar\omega|\beta|\gg E_{J}$, that can be obtained
for $|\beta| \geq 0.25$, the Hamiltonian in Eq.(\ref{HT0}) becomes
\begin{eqnarray}
H_{T}  & \cong & \hbar\omega
a^{\dag}a+i\frac{\hbar\omega}{2}\big[\big(\beta a-
\beta^{*}a^{\dag}\big) +
2i\frac{E_{z}}{\hbar\omega}\big]\sigma_{x}.\label{HT1}
\end{eqnarray}
Our Hamiltonian in Eq.(\ref{HT1}) becomes a Jaynes-Cummings type
Hamiltonian. The term $|\frac{\beta}{2}|^{2}$ was not taken into
account because it just represents an overall phase. Now we are
interested in the generation of Schr\"{o}dinger cat states ({\bf
SC}) in the above system. The time evolution of the state vector for
an initial state $|\psi(0)\rangle$ is
\begin{eqnarray}
|\psi(t) \rangle & = & T^{\dag} U_{T}(t)T|\psi(0)\rangle \nonumber
\\
& = & U(t)|\psi(0)\rangle ,\nonumber
\end{eqnarray}
where $U_{T}(t)=\exp{(-iH_{T}t/\hbar)}$ is the time evolution
operator in the transformation representation and $U(t) =
T^{\dag}U_{T}(t)T$ is the time evolution operator in the original
representation that is given by
\begin{eqnarray}
U(t)=T^{\dag}\exp\big\{-i\big[\omega
a^{\dag}a+i\frac{\omega}{2}\big[\big(\beta a- \beta^{*}a^{\dag}\big)
+ 2i\frac{E_{z}}{\hbar\omega}\big]\sigma_{x}\big]t\big\}T. \label{U}
\end{eqnarray}
Using the identity $e^{A+B}=e^{A}e^{B}e^{-[A,B]/2}$ the equation
(\ref{U}) may be written in the form
\begin{eqnarray}
U(t)&=& T^{\dag}e^{-i\omega
a^{\dag}at}e^{-\frac{\omega\beta}{2}(a^{\dag}-a)\sigma_{x}t}
e^{-\frac{\omega^{2}\beta}{4}(a^{\dag}+a)\sigma_{x}t^{2}}e^{i\frac{
E_{z}}{\hbar}\sigma_{x}t}T
\end{eqnarray}
where for simplicity we consider $\beta$ as real. After some
algebra, the time evolution operation in the original representation
becomes
\begin{equation}
U(t) = e^{-i\omega a^{\dag}at} \left(
\begin{array}{cc}
D(\tilde{\beta})e^{-i\frac{E_{z}}{\hbar}t} & 0 \\
0 & D^{\dag}(\tilde{\beta})e^{i\frac{E_{z}}{\hbar}t} \\
\end{array}
\right)
\end{equation}
where we define the parameter $\tilde{\beta} =
i\frac{\omega^{2}\beta}{4}t^{2}$ and $D(\tilde{\beta})$ is the
Glauber's displacement operator as defined above.

Considering the state vector as having the following initial
condition
\begin{equation}
|\psi(0) \rangle=|0\rangle\left[{1\over \sqrt{2}}\left(|e\rangle +
|g\rangle \right)\right],
\end{equation}
or the qubit prepared in a superposition states of $|e\rangle$ and
$|g\rangle$ and the field in a vacuum state $|0\rangle$,   the time
evolution of the state vector is given by
\begin{eqnarray}
|\psi(t) \rangle & = & U(t)|\psi(0)\rangle\nonumber\\
& = & \frac{1}{\sqrt{2}}
\left[e^{-i\frac{E_{z}}{\hbar}t}|e^{-i\omega
t}\tilde{\beta}\rangle|e\rangle +
e^{i\frac{E_{z}}{\hbar}t}|-e^{-i\omega
t}\tilde{\beta}\rangle|g\rangle\right]. \label{Psi1}
\end{eqnarray}

Now we rotate  the qubit so that $|e\rangle \rightarrow
\frac{1}{\sqrt{2}}(|e\rangle - |g\rangle)$ and $|g\rangle
\rightarrow \frac{1}{\sqrt{2}}(|e\rangle + |g\rangle)$. This
rotation is equivalent to applying the operator $R =
\frac{1}{\sqrt{2}}[1+(\sigma_{+}-\sigma_{+})]$ on the states
$|e\rangle$ and $|g\rangle$. Applying $R$ on the state given by
equation (\ref{Psi1}) a superposition of coherent states will be
obtained
\begin{equation}
R|\psi(t) \rangle = \frac{1}{\sqrt{2}} \left(\Phi_{+}|e\rangle -
\Phi_{-}|g\rangle\right)\label{SC}
\end{equation}
with the {\bf SC}
\begin{equation}
\Phi_{\pm} = \frac{1}{\sqrt{2}}
\left[e^{-i\frac{E_{z}}{\hbar}t}|e^{-i\omega
t}\tilde{\beta}\rangle|e\rangle \pm
e^{i\frac{E_{z}}{\hbar}t}|-e^{-i\omega
t}\tilde{\beta}\rangle|g\rangle\right].
\end{equation}

The result is an entangled state involving qubit and a cavity field.
If one measures the charge state (either in $|g\rangle$ or
$|e\rangle$), that action will collapse the $R|\psi(t)\rangle$ into
a {\bf SC} state $\Phi_{\pm}$. The form of Eq.(\ref{SC}) is very
similar to the {\bf SC} obtained in Ref. \cite{yu05}. But, in
contrast to that, we did not do use the condition $|\beta| \ll 1$.
In our scheme, as $|\beta|$ is large, and the value of the amplitude
of coherent states are proportional to $t^2$, the time for preparing
an observable {\bf SC} state is much shorter than that in other
schemes. The approach used here is similar to the approach used in
Ref. \cite{feng01} for the preparation of {\bf SC} with cold ions
beyond the Lamb-Dicke limit.

\section{Conclusion}
In conclusion, we have presented an approach for preparing {\bf SC}
states of mode of cavity field interacting with a superconducting
charge qubit. In contrast to others schemes we include nonlinear
effects. In general, approximations are made directly to the full
Hamiltonian in equation (\ref{H}) neglecting all higher orders of
$|\beta|$. In our scheme, we first apply an unitary transformation
to the Hamiltonian (\ref{H}) and make the relevant approximations
after performing the transformation. The result obtained holds for
any value of the parameter $\beta$. In the regime in which
$\hbar\omega |\beta| \gg E_{J}$, that can be obtained for$|\beta|
\ge 0.25$, the Hamiltonian becomes a Jaynes-Cummings type
Hamiltonian. Based on the measurement of charge states, we show that
{\bf SC} states of a single-mode cavity field can be generated.
Here, as $|\beta|$ is large, and the amplitude of the coherent
states are proportional to $t^2$, the time for preparing observable
{\bf SC} states is much shorter than in the linear regimes.

\ack D.S.F and M.C.N acknowledge the financial support from Conselho
Nacional de Desenvolvimento Cientifico e Tecnol\'ogico - CNPq
(150232/2012-8), Brazil.

\end{document}